\documentclass[useAMS,usenatbib]{mn2e}
\usepackage{graphicx}

\newcommand{\Porb}{\mbox{$P_{\mathrm{orb}}$}}
\newcommand{\Line}[3]{\Ion{#1}{#2}\,$\lambda$\,#3}

\newcommand{\Ion}[2]{#1{\,\scriptsize #2}}
\newcommand{\Twd}{\mbox{$T_{\mathrm{wd}}$}}
\newcommand{\Ha}{\mbox{${\mathrm{H\alpha}}$}}
\newcommand{\Hb}{\mbox{${\mathrm{H\beta}}$}}
\newcommand{\Hg}{\mbox{${\mathrm{H\gamma}}$}}
\newcommand{\id}{\mbox{$\mathrm{d^{-1}}$}}
\newcommand{\kms}{\mbox{$\mathrm{km\,s^{-1}}$}}
\newcommand{\es}{\mbox{$\mathrm{erg\;s^{-1}}$}}
\newcommand{\ecs}{\mbox{$\mathrm{erg\;cm^{-2}s^{-1}}$}}
\newcommand{\msy}{\mbox{$\mathrm{\Msun\,yr^{-1}}$}}
\newcommand{\cts}{\mbox{$\mathrm{cts\;s^{-1}}$}}

\newcommand{\Rwd}{\mbox{$R_{\mathrm{wd}}$}}
\newcommand{\Mwd}{\mbox{$M_{\mathrm{wd}}$}}
\newcommand{\Msun}{\mbox{$\mathrm{M}_{\odot}$}}

\title{A ZZ\,Ceti white dwarf in SDSS\,J133941.11+484727.5}

\author[B.T. G\"ansicke et al.]{
B.T. G\"ansicke$^1$,
P. Rodr{\'\i}guez-Gil$^2$, 
T.R. Marsh$^1$,
D. de Martino$^3$,
J. Nestoras$^4$,
P. Szkody$^5$,\newauthor
A. Aungwerojwit$^1$,
S.C.C. Barros$^1$,
M. Dillon$^1$,
S. Araujo-Betancor$^{2,6}$, 
M.J. Ar\'evalo$^{2,7}$,\newauthor
J. Casares$^2$, 
P.J. Groot$^8$, 
U. Kolb$^9$,
C. L\'azaro$^2$,
P. Hakala$^{10}$,
I.G. Mart{\'\i}nez-Pais$^{2,7}$,\newauthor
G. Nelemans$^8$,
G. Roelofs$^8$,
M.R. Schreiber$^{11,12}$,
E. van den Besselaar$^8$,
C. Zurita$^2$\\
$^1$ Department of Physics, University of Warwick, Coventry CV4 7AL,
UK \\
$^2$ Instituto de Astrof{\'\i}sica de Canarias, 38200 La Laguna, Tenerife, Spain\\
$^3$ INAF - Osservatorio di Capodimonte, Via Moiariello 16, 80131 Napoli, Italy\\
$^4$ Department of Physics, Section of Astrophysics, Astronomy \&
   Mechanics, University of Thessaloniki, 541 24 Thessaloniki, Greece \\
$^5$ Astronomy Department, University of Washington, Seattle, WA98195, USA\\
$^6$ Space Telescope Science Institute, 3700 San Martin Drive,
  Baltimore, MD21218, USA \\
$^7$ Departamento de Astrof{\'\i}sica, Universidad de La Laguna, E-38206 La
Laguna, Tenerife, Spain\\
$^8$ Department of Astrophysics, IMAPP, Radboud University Nijmegen,
  P.O. Box 9010, 6500 GL Nijmegen, The Netherlands\\
$^9$ Department of Physics and Astronomy, The Open University, Milton Keynes MK7 6AA, UK\\
$^{10}$ Observatory, University of Helsinki, PO Box 14, Helsinki, Finland\\
$^{11}$ Astrophysikalisches Institut Potsdam, An der Sternwarte 16, 14482 Potsdam, Germany\\
$^{12}$ Departamento de F{\'\i}sica y Meteorolog\'a, Facultad de Ciencias,
Universidad de Valpara{\'\i}so, Gran Breta\~na 644, Valpara{\'\i}so, Chile 
}

\begin{document}

\date{Accepted 2005. Received 2005; in original form 2005}

\pagerange{\pageref{firstpage}--\pageref{lastpage}} \pubyear{2005}

\maketitle

\label{firstpage}

\begin{abstract}
We present time-resolved spectroscopy and photometry of the
cataclysmic variable (CV) SDSS\,J133941.11+484727.5 (SDSS\,1339) which
has been discovered in the Sloan Digital Sky Survey Data Release
4. The orbital period determined from radial velocity studies is
82.524(24)\,min, close to the observed period minimum. The optical
spectrum of SDSS\,1339 is dominated to 90\% by emission from the white
dwarf. The spectrum can be successfully reproduced by a
three-component model (white dwarf, disc, secondary) with
$\Twd=12\,500$\,K for a fixed $\log g=8.0$, $d=170$\,pc, and a
spectral type of the secondary later than M8. The mass transfer rate
corresponding to the optical luminosity of the accretion disc is very
low, $\simeq1.7\times10^{-13}$\,\msy. Optical photometry reveals a
coherent variability at 641\,s with an amplitude of 0.025\,mag, which
we interpret as non-radial pulsations of the white dwarf. In addition,
a long-period photometric variation with a period of either 320\,min
or 344\,min and an amplitude of 0.025\,mag is detected, which bears no
apparent relation with the orbital period of the system. Similar
long-period photometric signals have been found in the CVs
SDSS\,J123813.73--033933.0, SDSS\,J204817.85--061044.8, GW\,Lib and
FS\,Aur, but so far no working model for this behaviour is available.
\end{abstract}


\section{Introduction}
A major impact of the Sloan Digital Sky Survey (SDSS) on the study of
cataclysmic variables (CVs) has been the discovery of more than 20
systems (by Data Release 4) in which the optical spectrum is dominated
by the white dwarf and there is no spectral signature of the donor star
\citep{szkodyetal02-2, szkodyetal03-2, szkodyetal04-1,
szkodyetal05-1}. These characteristics strongly suggest that those
CVs have very low mass transfer rates, low mass donors, and presumably
very short orbital periods. Hence, these systems resemble the old,
evolved CVs predicted by population models to exist in vast numbers
near the orbital period minimum \citep{kolb93-1,
howelletal97-1}. Whether or not their number agrees with the
theoretical models remains to be determined as their orbital periods
need to be measured, along with estimates of their distances and mass
transfer rates.

\begin{figure}
\centerline{\includegraphics[width=7cm]{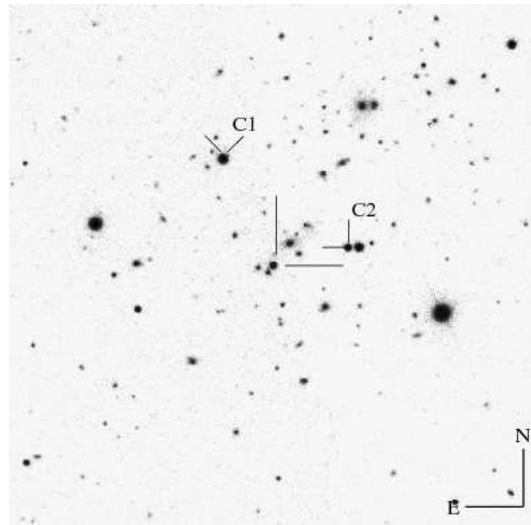}}
\caption{\label{f-fc} $5\arcmin\times5\arcmin$ finding chart of
SDSS\,1339 obtained from SDSS imaging data.  Primary and secondary
comparison stars for the CCD photometry are labelled `C1' ($g'=16.7$)
and `C2' ($g'=18.5$).}
\end{figure}

SDSS\,J133941.11+484727.5 (henceforth SDSS\,1339, Fig.\,\ref{f-fc}), a
CV discovered in Data Release 4 \citep{szkodyetal05-1}, is
characterised by a very strong contribution of the white dwarf in the
optical spectrum, and a clear absence of the TiO absorption bands typical
of a late-type main-sequence donor star in the red part of the
spectrum. Here we report detailed spectroscopic and photometric
follow-up observations of SDSS\,1339, which confirm a short orbital
period, a very low mass transfer rate, and identify a pulsating
ZZ\,Ceti-type white dwarf in this CV.

\section{Observations}
In 2004, the International Time Programme of the night-time telescopes
at the European Northern Observatory has been awarded for the study of
CVs identified in the SDSS, with the aim of improving our
understanding of compact binary evolution. The observations of
SDSS\,1339 were carried out as part of this project.

\subsection{Spectroscopy}
\begin{figure}
\includegraphics[angle=-90,width=\columnwidth]{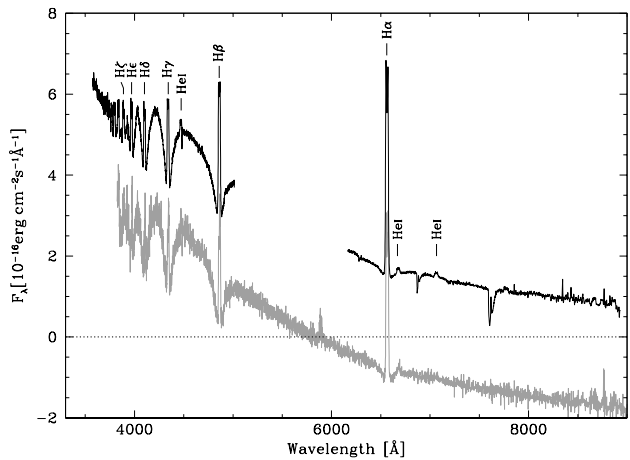}
\caption{\label{f-spec} Average WHT/ISIS spectra of SDSS\,1339 (black)
  and the SDSS discovery spectrum (gray, offset downwards by two flux
  units).  Noticeable emission lines are identified. The broad Balmer
  absorption lines reflect the dominant contribution of the white
  dwarf to the optical flux.}
\end{figure}

Time-resolved spectroscopy of SDSS\,1339 was obtained in January 2005
at the 4\,m William Herschel Telescope (WHT) on La Palma
(Table\,\ref{t-obslog}). The double-arm spectrograph ISIS was used
equipped with the R600B grating and a 4k$\times$2k pixel EEV detector
in the blue arm and the R316R grating and a 4.5k$\times$2k pixel
Marconi detector in the red arm. A 1.2\arcsec slit was used on both
arms, providing a spectral resolution of $\simeq0.9$\,\AA\ covering
the ranges $3600-5000$\,\AA\ and $6100-8900$\,\AA. The target
exposures were interleaved with arc lamp and flat-field exposures to
correct the wavelength scale for instrument flexure and to remove CCD
fringing in the red arm. Spectra of the spectroscopic standard star
Feige\,34 were obtained on each night immediately after the
observations of SDSS\,1339. The standard reduction of the spectra,
consisting of de-biasing, flat-fielding, optimal extraction and
wavelength and flux calibration, was carried out using \texttt{Figaro}
within the \texttt{STARLINK} suite and the \texttt{Pamela/Molly}
packages. The average of the blue and red spectra (Fig.\,\ref{f-spec})
is qualitatively similar to the SDSS spectrum \citep{szkodyetal05-1},
clearly dominated by the broad Balmer absorption lines from the white
dwarf photosphere. No noticeable signature of the companion star is
detected in the red end of the spectrum. The Balmer emission lines are
double-peaked, as typically observed in the quiescent spectra of
short-period dwarf novae with moderate orbital inclinations. The
emission lines of  \Ion{He}{I} are very weak.

\begin{table}
\begin{minipage}{\columnwidth}
\caption{\label{t-obslog}Log of the observations. Read-out times at the
WHT, TNG, and Kryoneri were $\sim40$\,s, $\sim10$\,s and $\sim5$\,s,
respectively.} \setlength{\tabcolsep}{1.1ex}
\begin{tabular}{llcccc}
\hline
Date & UT      &  Obs & Filter/Grism & Exp.(s)  & Frames \\
\hline
2005 Jan 02 & 06:04 -- 06:58 & WHT & R600B/R316R & 600 & 6 \\
2005 Jan 04 & 04:00 -- 06:54 & WHT & R600B/R316R & 400 & 25 \\
2005 Jan 05 & 04:08 -- 07:10 & WHT & R600B/R316R & 400 & 26 \\
2005 Jan 06 & 05:39 -- 06:51 & WHT & R600B/R316R & 400 & 6 \\
2005 Jan 07 & 04:40 -- 06:59 & WHT & R600B/R316R & 400 & 20 \\
2005 Apr 04 & 19:36 -- 00:24 & KY & Clear & 70 & 215 \\
2005 Apr 07 & 21:31 -- 03:03 & KY & Clear & 70 & 250 \\
2005 Apr 29 & 21:45 -- 01:43 & TNG & $g'$ & 20 & 432\\
\hline
\end{tabular}
\end{minipage}
\end{table}

\subsection{Photometry}
\begin{figure}
\includegraphics[angle=-90,width=\columnwidth]{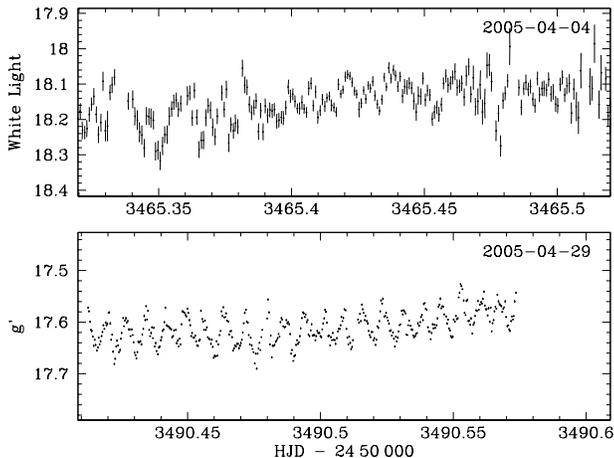}
\caption{\label{f-lc} Kryoneri (top) and TNG (bottom) light curves of
SDSS\,1339. The average brightness of the TNG data agrees with the
$g'$-band measurement from SDSS. The Kryoneri data were obtained in
white light, the difference in mean magnitude is likely due to colour
differences between the target and the comparison star.}
\end{figure}

Filterless CCD photometric time-series of SDSS\,1339 were obtained in
April 2005 at the 1.2\,m Kryoneri telescope using a Photometrics SI-502
$516\times516$ pixel camera (Table\,\ref{t-obslog}). The object images
were corrected for bias, dark current and flat field structures within
\textsc{MIDAS}.  Subsequently, the \textsc{Sextractor} package
\citep{bertin+arnouts96-1} was used to perform aperture photometry on
the processed images. The differential light curves of SDSS\,1339 were
computed relative to the comparison star `C1' ($g'=16.7$), and a second
comparison star `C2' ($g'=18.5$) was used to verify that `C1' is not
variable and to monitor changes in the observing conditions.  The
Kryoneri light curves (Fig.\,\ref{f-lc}, top panel) gave some evidence
for short-term variability, and power spectra computed from both
nights contained a signal near 135\,\id\ ($\simeq10.7$\,min). Prompted
by this discovery, we obtained additional $g'$-band CCD photometry at
the 3.6\,m Telescopio Nazionale Galileo (TNG) using DOLORES equipped
with a Loral $2\mathrm{k}\times2\mathrm{k}$ pixel CCD. The CCD was
binned $2\times2$ and windowed in order to reduce the readout
time. The TNG data were reduced in the same manner as described above,
and the light curve (Fig.\,\ref{f-lc}, lower panel) clearly confirms the
presence of photometric variability at a period of $\sim10$\,min.

\section{System Parameters}

\subsection{The orbital period}
\label{s-porb}
\begin{figure}
\includegraphics[width=\columnwidth]{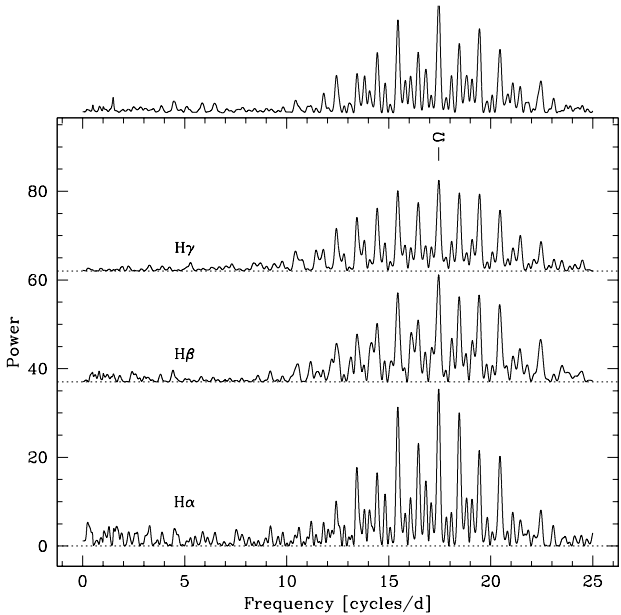}
\caption{\label{f-rvscargle} Main panel: Scargle periodograms
  computed from the radial velocity variations of the \Ha\ (bottom
  curve), \Hb\ (middle curve) and \Hg\ (top curve) line wings in
  SDSS\,1339. The implied orbital period is $\Porb=82.524$\,min.  Top
  Panel: Scargle periodogram of a set of fake radial velocity data
  computed from a sine wave with $P=82.524$\,min with the same
  temporal sampling as the observed data.}
\end{figure}

\begin{figure}
\includegraphics[angle=-90,width=\columnwidth]{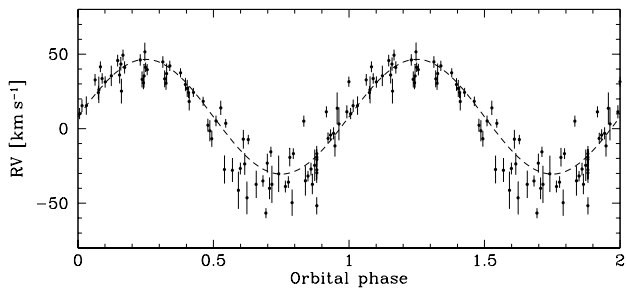}
\caption{\label{f-rv} Radial velocity (RV) variation of \Ha\
  measured from the time-resolved WHT spectroscopy of SDSS\,1339
  (Table\,\ref{t-obslog}), folded over the orbital period. The
  data are repeated over two cycles for clarity.}
\end{figure}

In order to determine the orbital period of SDSS\,1339 from a radial
velocity study, we applied heliocentric corrections to all spectra and
binned them onto a homogeneous wavelength scale. The radial velocity
variation of the \Ha\ emission line wings were then measured using the
double-Gaussian convolution technique described by
\citet{schneider+young80-1} with a full width at half maximum of the
individual Gaussians of 200\,\kms\ and a separation of 1500\,\kms. The
strongest signal in the \citet{scargle82-1} periodogram computed from
the radial velocity data (Fig.\,\ref{f-rvscargle}) is detected at a
period of 82.524(24)\,min, where the error has been determined by
fitting a sine wave to the radial velocity data. A fake data set
constructed from a sine wave sampled at the same times of the
observations, with a period of 82.524\,min, and amplitude and error
distribution similar to the observed radial velocities results in a
nearly identical periodogram, confirming that the relatively complex
structure of the periodogram is due to the temporal sampling of our
spectroscopic data.  We interpret this period as the orbital period of
SDSS\,1339. Folding the \Ha\ radial velocity measurements over the
orbital period produces a quasi-sinusoidal curve. A sine fit gives an
amplitude of $39\pm2$\,\kms\ and a $\gamma$-velocity of $8\pm1$\,\kms
(Fig.\,\ref{f-rv}). We determine the orbital ephemeris
\begin{equation}
\label{e-ephemeris}
T_0 = \mathrm{HJD}\,2453372.74861(37) + 0.057309(17)\times E
\end{equation}
from the sine fit to the radial velocity measurements, where $T_0$ is
the time of inferior conjunction of the secondary star if the radial
velocity variation of the line wings traces the motion of the white
dwarf. We caution, however, this interpretation, as the signal in the
line wings is likely to be contaminated to some extent by the bright
spot where the mass transfer stream from the secondary impacts the
accretion disc.  The same method was applied to the emission lines of
\Hb\ and \Hg, and resulted in consistent values for \Porb, though with
larger errors as a consequence of the lower signal-to-noise ratio and
the larger disturbance due to the white dwarf absorption lines in
these lines compared to \Ha\ (Fig.\,\ref{f-rvscargle}). Sine-fits to
the radial velocities of \Hb\ and \Hg\ give amplitudes of
$47\pm3$\,\kms\ and $60\pm5$\,\kms, respectively and
$\gamma$-velocities of $1\pm2$\,\kms\ and $8\pm3$\,\kms, respectively.

\begin{figure*}
\includegraphics[width=16cm]{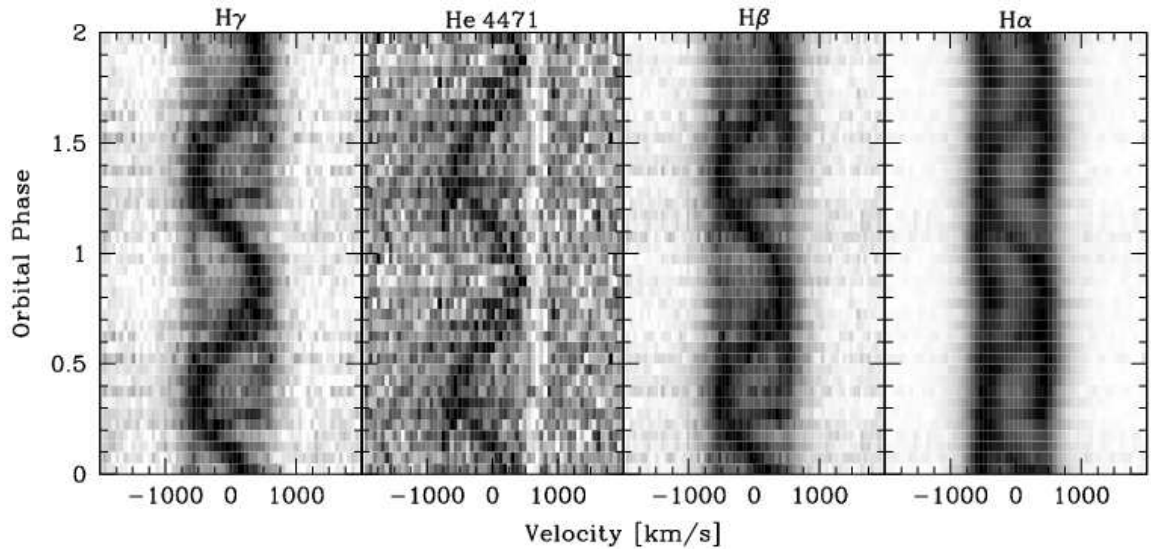}
\caption{\label{f-trail} Trailed spectrograms of \Hg,
    \Line{He}{I}{4471}, \Hb, and \Ha\ folded on the orbital period of
    82.524\,min. The data are repeated over two cycles for clarity.}
\end{figure*}

Figure\,\ref{f-trail} shows trailed spectrograms for \Hg,
\Line{He}{I}{4471}, \Hb, and \Ha\ folded on 20 phase bins, using our
ephemeris. The double-peaked profiles of \Ha, \Hb\ and \Hg\ are
apparent, with a half-separation of $\sim500$\,\kms. There is a
prominent S-wave whose semi-amplitude is consistent with this value,
indicating an origin in the outer edge of the disc, very likely the
bright spot. The half-separation of the double-peaked profiles
together with the absence of eclipses in the light curves suggest an
intermediate inclination. The absence of a photometric modulation that
could be ascribed to the bright spot suggests that the bright spot
contributes primarily in the Balmer emission lines. From the trailed
spectra, it appears that the relative contribution of the bright spot
increases for the higher members of the Balmer series, implying a
stronger Balmer decrement in the disc compared to the bright spot.

\subsection{Additional system parameters}
We have modelled the average WHT spectrum of SDSS\,1339 as the sum of
three individual components: the white dwarf, the accretion disc, and
the donor star. The white dwarf is represented by synthetic spectra
computed with \citet{hubeny+lanz95-1}'s TLUSTY/SYNSPEC codes. We
generated a grid of pure-hydrogen models covering the range
8000\,K--20\,000\,K with the surface gravity fixed to $\log g=8.0$
($\Mwd\simeq0.6$\,\Msun). For the white dwarf radius, we assume
$\Rwd=8.7\times10^8$\,cm following the \citet{hamada+salpeter61-1}
mass-radius relation for zero-temperature carbon-oxygen white dwarfs.
The accretion disc is represented by the emission of an isothermal and
isobaric hydrogen slab, following the description by
\citet{gaensickeetal97-1, gaensickeetal99-1}. The secondary star is
represented by observed templates covering M0.5 to M9 from
\citet{beuermannetal98-1} and L0 to L8 from
\citet{kirkpatricketal99-1} and \citet{kirkpatricketal00-1}. We fix
the radius of the secondary star to $R_2=8.6\times10^9$\,cm,
corresponding to a $M_2=0.08$\,\Msun\ donor star at the low-mass end
of the main sequence (the Roche-lobe radius of the secondary for this
choice of $M_2$ at the given orbital period). $R_2$ obviously depends
on the choice of $M_2$ (and fairly little on $\Mwd$, which was fixed
to 0.6\,\Msun, as stated above), but given that the spectral type-mass
relation for CV donors at such low masses is undetermined we consider
this a justified simplification.

Our approach is a forward-modelling rather than a fitting, and
proceeds as follows. Free parameters are the white dwarf temperature
\Twd, the distance $d$ to SDSS\,1339, the temperature $T_\mathrm{d}$
and column density $\Sigma_\mathrm{d}$ of the disc, and the spectral
type of the secondary star Sp(2). As a first step, the disc spectrum
for a given choice of ($T_\mathrm{d}$, $\Sigma_\mathrm{d}$) is
normalised to the observed flux of \Ha. The second step consists of
choosing \Twd\ and adjusting the flux scaling factor of the model
spectrum in a way that the sum of disc plus white dwarf fits the
observed flux level in the spectrum from the blue arm. At this stage,
the distance is determined by the knowledge of \Rwd\ and the flux
scaling factor, and the spectrum of a secondary star of spectral type
Sp(2) is added to the model, scaled appropriately for $R_2$ and
$d$. This procedure is optimised until (a) the white dwarf model
adequately reproduces the observed Balmer absorption lines, (b) the
disc emission reproduces the observed emission line flux ratios, (c)
the flux contribution from the secondary is sufficiently low to be
consistent with the non-detection of molecular absorption bands in the
red arm spectrum, and (d) the overall slope of the observed continuum
is reproduced. The best-matching set of parameters is found to be
$\Twd=12\,500$\,K, $T_\mathrm{d}=6600$\,K,
$\Sigma_\mathrm{d}=1.7\times10^{-2}\,\mathrm{g\,cm^{-2}}$,
$d=170$\,pc, and Sp(2) later than M8. The luminosity of the accretion
disc is $\simeq10^{30}$\,\es, less than 10\% of the white dwarf
luminosity. For an assumed white dwarf mass of 0.6\,\Msun, this
luminosity corresponds to to an accretion rate of $\dot
M\simeq1.7\times10^{-13}$\,\msy. A caveat to white dwarf temperature
and distance determination is the unknown mass of the
primary. Masses of single white dwarfs
\citep[e.g.][]{koesteretal79-1, bergeronetal92-1, liebertetal05-1} are
predominantly clustered sharply around $\sim0.6$\,\Msun. The standard
method used for single white dwarfs, modelling the Balmer absorption
lines, is not available for CV white dwarfs due to contamination by
the accretion disc/stream and the secondary star, and white dwarf mass
estimates have to be determined from radial velocity studies, eclipse
timing, or ultraviolet spectral modelling. Consequently, the number of
CVs with undisputed mass measurements is very small. The secular
evolution of CV white dwarf masses depends on the ratio of accreted
material to material ejected during classical novae eruptions, and the exact
details are not very well established \citep{yaronetal05-1}. In the
light of the uncertain mass of the white dwarf in SDSS\,1339, we have
therefore repeated the above analysis assuming a white dwarf mass
higher (lower) by 0.3\,\Msun\ and find a white dwarf effective
temperature higher (lower) by $\simeq500-1000$\,K and a distance lower
(larger) by $\simeq35$\,pc.

If we assume that the disc emission is all what there is in terms of
accretion luminosity, the implied accretion rate is very low, in fact,
much lower than mass transfer rates predicted from angular momentum
loss via gravitational radiation, $\simeq3.5\times10^{-11}$\,\msy\ for
a 0.6\,\Msun\ white dwarf. Some additional accretion luminosity may be
released in X-rays. The ROSAT All Sky Survey \citep{vogesetal00-1}
contains a faint source, 1RXS\,J133941.9+484844, with
$1.23\pm0.56\times10^{-2}$\,\cts, that within its large positional
uncertainty (77.4\arcsec) coincides with SDSS\,1339. A pointed
observation of that region led to the detection of
2RXP\,J133938.5+484722, with $1.5\times10^{-2}$\,\cts. No error on the
position and count rate are given in the \textit{The Second ROSAT
Source Catalog of Pointed Observations}.  It is possible that both
X-ray sources are identical, and that SDSS\,1339 is the optical
counterpart.  If the X-rays detected by ROSAT are indeed associated
with SDSS\,1339, the observed PSPC count rate corresponds to an
unabsorbed bolometric flux of of $\sim2\times10^{-13}$\,\ecs, assuming
a thermal Bremsstrahlung spectrum with $2\,\mathrm{keV}\la kT\la5$\,keV, as
typically observed in short-period CVs, and a neutral hydrogen column
density of $N_\mathrm{H}=1.1\times10^{20}\,\mathrm{cm^{-2}}$. For
$d=170$\,pc, the implied X-ray luminosity would be $\sim10^{30}$\,\es,
very similar to the optical accretion luminosity derived above, and
comparable to the X-ray emission of the well-studied minimum period CV
WZ\,Sge \citep{mukai+patterson04-1}.  A deeper X-ray observations is
clearly desirable to test whether SDSS\,1339 is the source of the
X-rays detected by ROSAT.

In summary, in SDSS\,1339 the accretion disc contributes less than
10\% of the optical light, which is very low compared to other
short-period CVs, and may have an X-ray flux comparable to that of
WZ\,Sge, suggesting a very low accretion rate. A low mass transfer
rate from the secondary is also supported by the absence of the
signature from a bright spot in the light curve (Fig.\,\ref{f-lc},
see also Fig.\,\ref{f-scargle}). The current data suggest
that the mass transfer from the secondary is lower than that predicted
by gravitational radiation. Following \citet{townsley+bildsten03-1},
the white dwarf temperature can be used as an estimate of the secular
mean of the accretion rate, and it appears that the white dwarf in
SDSS\,1339 is not unusually cold compared to other CVs near the
orbital minimum. In fact, its temperature is well in line with the
predictions for $\dot M$ from gravitational radiation. A possible
solution to the discrepancy between the accretion rate estimated from
the optical emission of the accretion disc (and the possible X-ray
component) is that the system is currently accreting below its secular
mean rate.

\begin{figure}
\includegraphics[angle=-90,width=\columnwidth]{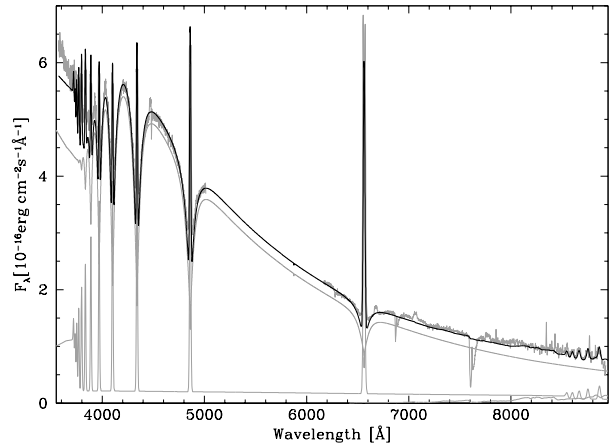}
\caption{\label{f-opt_fit} Three-component model (white dwarf, optically
  thin disc, secondary star) of the average WHT spectrum of
  SDSS\,1339. The observed data as well as the three individual
  components are plotted as gray lines, the summed model as a black
  line. The parameters are $\Twd=12\,500$\,K (assuming $\log g=8.0$),
  $T_\mathrm{d}=6600$\,K,
  $\Sigma_\mathrm{d}=1.7\times10^{-2}\,\mathrm{g\,cm^{-2}}$,
  Sp(2)\,=\,M8, and $d=170$\,pc.}
\end{figure}

\section{A ZZ\,Ceti-type white dwarf}
\begin{figure}
\includegraphics[angle=-90,width=\columnwidth]{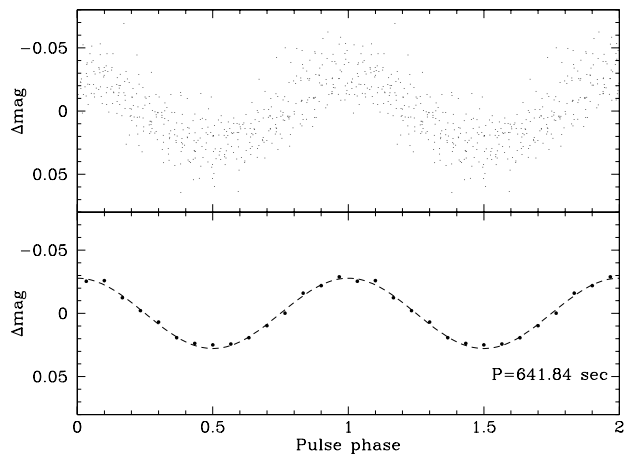}
\caption{\label{f-spinfold} Pulse-folded light curves of SDSS\,1339,
  assuming $P=641.84$\,s. The long-period 344.32\,min modulation
  (Sect.\,\ref{s-longperiod}) has been removed. 
  Top panel: all data points from the TNG. Bottom panel: TNG data binned
  into 15 phase slots. Plotted as dashed line is a sine fit to the
  binned and folded data. The
  data are repeated over two cycles for clarity.}
\end{figure}

The power spectra computed from the Kryoneri and TNG photometric data
consistently contain a strong signal at 134.61\,\id\
(Fig.\,\ref{f-scargle}).  Given the estimated temperature of the white
dwarf, the most obvious explanation for this signal are non-radial
pulsations with a period of $P=641.84$\,s. The amplitude of the
pulsation is $\simeq0.025$\,mag, which is well within the range
observed in single ZZ\,Ceti white dwarfs
\citep{winget98-1}. Figure\,\ref{f-spinfold} shows the TNG photometry
folded over the pulse period of 641.84\,s, after subtracting a
long-period photometric modulation with $P=344.32$\,min (see
Sect.\,\ref{s-longperiod}). The detection of pulsations makes
SDSS\,1339 only the eighth pulsating white dwarf in a CV.

The fact that only one pulsation mode is detected above a 3-$\sigma$
threshold raises the question whether other mechanisms than non-radial
pulsations could cause the observed variability. In principle, the
white dwarf spin could account for a stable clock in photometric data,
as seen in intermediate polars. However, in intermediate polars,
multiple optical modulations are observed, usually at the white dwarf
rotational period, the orbital period and sideband periods. In these
systems the spin modulation stems the reprocessing of X-rays in the
magnetically confined accretion flow onto the WD while the orbital and
the beat periodicities are due to reprocessing in the accretion disc
and the bright spot. However, in SDSS\,1339 we do not detect a
significant signal at the orbital period, nor any beat
signal. Furthermore, the optical spectrum of SDSS\,1339 in no way
resembles that of any confirmed intermediate polar, as it has
literally no \Line{He}{II}{4686} emission, and not a single
intermediate polar is known whose optical emission is dominated by the
white dwarf. The non-detection of Zeeman splitting in the Balmer
lines, specifically the higher members, limits the possible field
strength of the white dwarf to $B<1$\,MG. Finally, as discussed above,
SDSS\,1339 is not a prominent X-ray source.

\section{A long non-\Porb\ photometric periodicity}
\label{s-longperiod}
\begin{figure}
\includegraphics[width=\columnwidth]{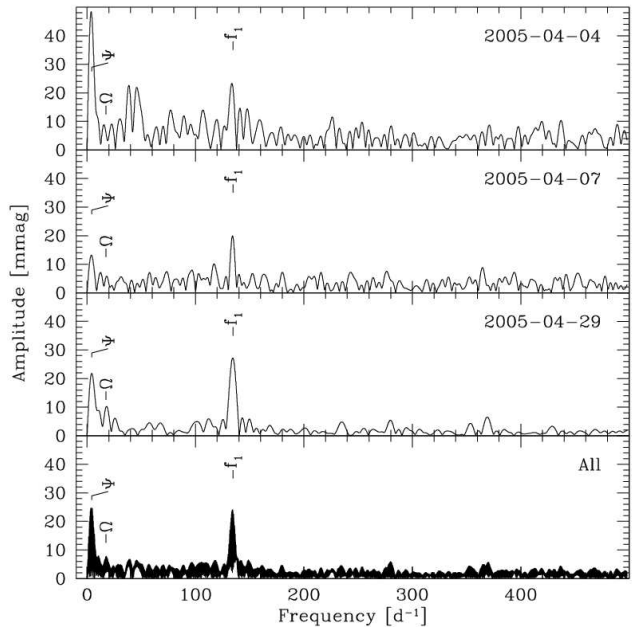}
\caption{\label{f-scargle} Power spectra computed from the three
  individual nights of photometry obtained at Kryoneri Observatory and
  with the TNG, and from the combined data. $f_1$ and $\Psi$ indicate
  the frequencies of the white dwarf pulsation (641.84\,s) and a
  photometric modulation of unknown nature (319.95\,min or
  344.32\,min). No significant signal is detected in the power spectra
  at the orbital period $\Omega$ (82.524\,min).}
\end{figure}

\begin{figure}
\includegraphics[angle=-90,width=\columnwidth]{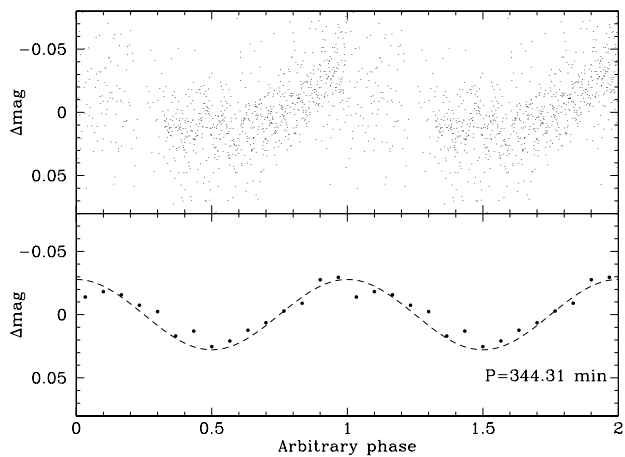}
\caption{\label{f-longfold} All photometric data (Kryoneri \& TNG)
  folded over the photometric period of 344.32\,min. The 641.84\,s
  pulse period has been removed. Top panel: all
  individual data points.  Bottom panel: Data binned into 15 phase
  slots. Plotted as a dashed line is a sine fit to the binned and folded
  data. The
  data are repeated over two cycles for clarity.}
\end{figure}

Inspection of the power spectra (Fig.\,\ref{f-scargle}) reveals a
low-frequency signal of variable amplitude in all three nights. The
power spectrum of the combined data contains two signals of nearly
identical amplitude at 4.50\,\id\ and 4.18\,\id, with the latter one
being the stronger peak. Removing the photometric data with the
641.84\,s pulse signal and fitting them to a sine wave results in two
possible periods, 319.95(5)\,min or
344.32(7)\,min. Figure\,\ref{f-longfold} shows the pulse-removed
photometry folded over 344.32\,min. The average amplitude of this
modulation is 0.025\,mag, but the power spectra indicate that the
amplitude  is rather variable on a time scale of days. No equivalent
long-period variability is detected in the radial velocity variations
(Sect.\,\ref{s-porb}).

Long-period photometric modulations with periods in the range 7--12\,h
have been reported for SDSS\,J123813.73--033933.0 and
SDSS\,J204817.85--061044.8 \citep{zharikovetal05-1, woudtetal05-1},
which are both short-period CVs with white-dwarf dominated optical
spectra.  A similar phenomenon is also well-documented in GW\,Lib,
where a 125.4\,min photometric modulation has been detected with an
amplitude of $\simeq0.05$\,mag on several occasions
\citep{woudt+warner02-2}. In FS\,Aur, a 205.5\,min photometric
modulation with a 0.24\,mag amplitude was detected by
\citet{tovmassianetal03-1}.  In both systems, the photometric
modulations occur on periods substantially longer than the orbital
periods (76.9\,min and 85.7\,min,
respectively). \citet{tovmassianetal03-1} invoke the precession of a
rapidly rotating white dwarf as a possible explanation for the
long-period signal found in FS\,Aur. To date, this hypothesis could
not be confirmed, and no other plausible model has been suggested.

The puzzle of long-period signals in short-period CVs is exacerbated
by the case of HS\,2331+3905, where a \textit{spectroscopic}
(i.e. radial velocity) periodicity of $\sim3.5$\,h is found that is in
no way associated with the 81.1\,min orbital period of the system, and
that does not have any photometric equivalent
\citep{araujo-betancoretal05-1}.  Whereas there is mounting evidence
that variability on periods much longer than the orbital period is
fairly common among CVs near the minimum period, the origin of this
phenomenon is unclear.

\begin{table*}
\caption{\label{t-cvzz}Properties of pulsating white dwarfs in
  cataclysmic variables}
\begin{tabular}{lcccccll}
\hline
System & \Porb\ [min] & Mag &  periods [s] & Outbursts & 
  \Twd\ [K] / $\log g$ & Ref \\
\hline
PQ\,And       & 80.7/78.5$^p$ & $V\simeq19.1$   & 1263, 634 & 1938, 1967, 1988 &
   12\,000/7.7$^a$ & 1,2,3 \\ 
GW\,Lib       & 76.8$^s$ & $V\simeq16.7^b$   & 650, 370, 230 & 1983 & 
   11\,000/8.0$^c$ 14\,700/8.0$^d$ & 4,5,6,7,8\\
HS\,2331+3905 & 81.1$^{pe}$ & $V\simeq16.5$ & 310, 336, 419$^e$ & - &
   10\,500/8.0$^f$ & 9 \\
RE\,J1255+266 & 119.4$^p$ & $V\simeq19.2$ & 668, 1236, 1344 & 1994 & 
   36\,480/9.0$^g$ 11\,000--15\,000$^h$ & 10,11,12\\
SDSS\,J013132.39-090122.3 & 98:$^s$ & $g'\simeq18.3$ & 260, 335, 595 & - & 
   - & 13,14 \\
SDSS\,133941.11+484727.5 & 82.5$^s$ & $g'\simeq17.6$ & 642 & - & 
  12\,500/8.0$^c$  & 15, this paper \\
SDSS\,J161033.64-010223.3 & 80.5$^p$ & $g'\simeq19.1$ & 607, 345, 304, 221 & - &
   - & 15,16  \\
SDSS\,J220553.98+115553.7 & - & $g'\simeq20.1$ & 330, 475, 575 & - & 
   - & 13,14\\
\hline
\end{tabular}
\begin{minipage}{\textwidth}
%
$^1$\citet{schwarzetal04-1},
$^2$\citet{pattersonetal05-1},
$^3$\citet{vanlandinghametal05-1},
%
$^4$\citet{szkodyetal00-1},
$^5$\citet{szkodyetal02-4},
$^6$\citet{thorstensenetal02-3}, 
$^7$\citet{thorstensen03-1},
$^8$\citet{vanzyletal04-1},
%
$^9$\citet{araujo-betancoretal05-1},
%
$^{10}$\citet{watsonetal96-1},
$^{11}$\citet{wheatleyetal00-1},
$^{12}$\citet{pattersonetal05-2},
%
$^{13}$\citet{szkodyetal03-2},
$^{14}$\citet{warner+woudt04-1},
$^{15}$\citet{szkodyetal05-1},
%
$^{16}$\citet{szkodyetal02-2},
$^{17}$\citet{woudt+warner04-1}. 
%
%
\newline
: Uncertain estimate. $^p$ Photometric period. $^{pe}$ Photometric period, eclipsing. $^s$ Spectroscopic period.
$^a$ Fit to the  Balmer lines.
$^b$ \citet{downesetal01-1} list a pre-outburst photometric magnitude
of $\simeq18.5$, however, most published pre- and post-outburst
photometry suggests $V\simeq16.7$ (USNO, DENIS,
\citealt{thorstensen03-1}). 
$^c$ Fit to the  Balmer lines with fixed $\log g$.
$^d$ Fit to far-ultraviolet spectra with fixed $\log g$.
$^e$ Very complex power spectrum.
$^f$ Fit to far-ultraviolet/optical spectrum with fixed $\log g$.
$^g$ Fit to the Balmer lines, using a model grid with
   $20\,000\,\mathrm{K}\le\Twd\le40\,000\,\mathrm{K}$. 
$^h$ Estimated from $UBVI$ colours.
\end{minipage}
\end{table*}

\section{Discussion}
Asteroseismology of single white dwarfs is a well-established field,
with close to 100 ZZ\,Ceti stars known (more than half of which
were discovered in the SDSS, \citealt{mukadametal04-1,
mullallyetal05-1}). \citet{winget98-1} and \citet{clemens93-1} note a
substantial diversity of the single pulsating ZZ\,Ceti stars as a
function of their temperature. The hot ($\Twd\simeq12\,000$\,K)
pulsators have a relatively small number of modes, typically in the
range 100\,--300\,s, with a high degree of stability both in
frequencies and amplitudes. In contrast to this, the cooler
($\Twd\simeq11\,000$\,K) pulsators are characterised by longer
periods in the range 600--1000\,s, larger amplitudes, more modes, and
what appears to be unstable amplitudes of the individual
modes. Overall, the power spectra of the cooler pulsators are much
more challenging to interpret than those of the hotter ones.

Since the discovery of non-radial pulsations of the white dwarf in
GW\,Lib \citep{vanzyletal04-1}, much effort has been invested in
identifying additional CV white dwarf pulsators
\citep[e.g.][]{woudt+warner04-1, araujo-betancoretal05-1,
pattersonetal05-2} and in theoretical modelling of accreting white
dwarf pulsators \citep{townsleyetal04-1}. The scientific potential of
asteroseismology in CVs is enormous, as it could allow accurate
measurements of white dwarf masses, envelope masses, rotation rates,
and magnetic fields~--~important parameters for understanding the
structure and evolution of CVs which are so far difficult or not at
all measurable. From the current roster of known CV white dwarf
pulsators given in Table\,\ref{t-cvzz}, it is clear that there is
still a long way to go before achieving these goals. 

An observational problem is that the CV white dwarf pulsators are all
much fainter than the field ZZ\,Ceti stars, in fact, hardly any single
white dwarf pulsator with $V>19$ has been studied. The need to detect
low-amplitude variability in these objects requires photometry
obtained at large-aperture telescopes, where it is difficult to obtain
observing runs long enough to determine accurate pulsation
frequencies. The largest photometric data sets have been published for
the two brightest CV pulsators, GW\,Lib and HS\,2331+3905
\citep{vanzyletal04-1, araujo-betancoretal05-1}. Both systems exhibit
large variations in their power spectra on time scales of days to
months, and no detailed mode identification has been achieved so far. 

A problem intrinsic to the nature of accreting sources is the
determination of the white dwarf temperature. Using optical
wavelengths alone gives rise to large systematic uncertainties as any
fit to the Balmer absorption lines is subject to an unknown
contribution from the accretion disc (and possibly the bright
spot). Given how narrow the ZZ\,Ceti instability strip of single white
dwarfs is in terms of temperature, there is little hope in empirically
defining the equivalent for CV white dwarfs from optical data alone.
A substantial improvement comes from far-ultraviolet (FUV) data, as
the white dwarf dominates in this wavelength range, and both the
Ly$\alpha$ absorption profile as well as the FUV to optical spectral
energy distribution can be used in estimating the white dwarf
temperature. Interestingly enough, again GW\,Lib and HS\,2331+3905 are
the only systems with accurate white dwarf temperatures, and the FUV
data for GW\,Lib indicate that parameters might be very different in
CV pulsators compared to field ZZ\,Ceti stars: the FUV-determined
temperature is 14\,700\,K for an assumed $\log g=8.0$\footnote{In
contrast to single white dwarfs, it is very difficult if not
impossible to break the degeneracy between \Twd\ and $\log g$ that
occurs in spectral modelling. The reason is that the Balmer lines are
strongly contaminated by the accretion disc, and that the FUV
observations provided by \textit{HST}/STIS cover only the red wing of
Ly$\alpha$.}, well outside the instability strip of single ZZ\,Ceti
stars \citep{szkodyetal02-4}. \citet{szkodyetal02-4} note that the FUV
spectrum is better fitted with a two-temperature white dwarf model,
where the lower temperature gets fairly close to the hot edge of the
ZZ\,Ceti instability strip. However, the physical origin of such an
inhomogeneous temperature distribution over the white dwarf is not
clear. HS\,2331+3905 has a temperature just close to the cold edge of
the ZZ\,Ceti instability strip, which is consistent with its very
complex power spectrum
\citep{araujo-betancoretal05-1}. 

Currently, all known CV white dwarf pulsators have orbital periods
very close to the orbital period minimum (Table\,\ref{t-cvzz}), where
accretion rates are sufficiently low to correspond to white dwarf
effective temperatures close to the ZZ\,Ceti instability strip. The
immediate task in improving our understanding of the pulsations in
accreting white dwarfs is now to measure accurate effective
temperatures for all white-dwarf dominated CVs, establish whether or
not they are pulsating, and determine their pulsation
frequencies. Only once the pulsation modes have been identified,
asteroseismology may reveal details about the structure of these
stars.

\section{Conclusions}
We have determined the orbital period of SDSS\,1339 to be 82.524\,min
from radial velocity studies. The optical spectrum is dominated by the
white dwarf, and can be successfully modelled with a (12\,500\,K,
$\log g=8.0$) synthetic spectrum. The contribution of the accretion
disc to the observed optical flux is less than 10\%, corresponding to
a very low optical accretion luminosity of
$\simeq1.7\times10^{-13}$\,\msy.  CCD photometry of SDSS\,1339 reveals
the presence of variability with a stable 641\,s period and an
amplitude of $\simeq0.025$\,mag, which we interpret as non-radial
pulsations of the white dwarf. Further photometric work is encouraged
to probe for additional pulsation modes.  Similar to several other
short-period CVs, SDSS\,1339 displays an apparently coherent
photometric variability on time scales much longer than the orbital
period. The nature of this variability is not understood.

\section*{Acknowledgements}
BTG and TRM were supported by a PPARC Advanced Fellowship and a
PPARC Senior Fellowship.
DdM acknowledges funding from the Italian Ministry of University and
Research (MIUR).
PS acknowledges some support from NSF grant AST 02-05875.
AA thanks the Royal Society for generous funding.
PJG, GR, EvdB are supported by NWO-VIDI grant 639.042.201 and GN is
supported by NWO-VENI grant 639.041.405.
MRSK thanks for supported by the Deutsches Zentrum f\"ur Luft-
und Raumfahrt (DLR) GmbH under contract No. FKZ 50 OR 0404.
Based in part 
on observations made with the William Herschel Telescope, which is
operated on the island of La Palma by the Isaac Newton Group in the
Spanish Observatorio del Roque de los Muchachos of the Instituto de
Astrof{\'\i}sica de Canarias (IAC);
on observations made with the Telescopio Nazionale Galileo 
operated on the island of La Palma by the Centro Galileo Galilei of
the INAF (Istituto Nazionale di Astrofisica) at the Spanish
Observatorio del Roque de los Muchachos of the IAC;
and on observations made at the 1.2\,m telescope, located at Kryoneri
Korinthias, and owned by the National Observatory of Athens, Greece.
The WHT and TNG data were obtained as part of the 2004 International
Time Programme of the night-time telescopes at the European Northern
Observatory.
We thank the referee for helpful comments.


\begin{thebibliography}{48}
\expandafter\ifx\csname natexlab\endcsname\relax\def\natexlab#1{#1}\fi

\bibitem[{{Araujo-Betancor} et~al.(2005)}]{araujo-betancoretal05-1}
{Araujo-Betancor}, S., et~al., 2005, A\&A, 430, 629

\bibitem[{{Bergeron} et~al.(1992){Bergeron}, {Saffer}, \&
  {Liebert}}]{bergeronetal92-1}
{Bergeron}, P., {Saffer}, R.~A., {Liebert}, J., 1992, ApJ, 394, 228

\bibitem[{{Bertin} \& {Arnouts}(1996)}]{bertin+arnouts96-1}
{Bertin}, E., {Arnouts}, S., 1996, A\&AS, 117, 393

\bibitem[{{Beuermann} et~al.(1998){Beuermann}, {Baraffe}, {Kolb}, \&
  {Weichhold}}]{beuermannetal98-1}
{Beuermann}, K., {Baraffe}, I., {Kolb}, U., {Weichhold}, M., 1998, A\&A, 339,
  518

\bibitem[{{Clemens}(1993)}]{clemens93-1}
{Clemens}, J.~C., 1993, Baltic Astronomy, 2, 407

\bibitem[{{Downes} et~al.(2001){Downes}, {Webbink}, {Shara}, {Ritter}, {Kolb},
  \& {Duerbeck}}]{downesetal01-1}
{Downes}, R.~A., {Webbink}, R.~F., {Shara}, M.~M., {Ritter}, H., {Kolb}, U.,
  {Duerbeck}, H.~W., 2001, PASP, 113, 764

\bibitem[{{G\"ansicke} et~al.(1997){G\"ansicke}, {Beuermann}, \&
  {Thomas}}]{gaensickeetal97-1}
{G\"ansicke}, B.~T., {Beuermann}, K., {Thomas}, H.~C., 1997, MNRAS, 289, 388

\bibitem[{{G\"ansicke} et~al.(1999){G\"ansicke}, {Sion}, {Beuermann}, {Fabian},
  {Cheng}, \& {Krautter}}]{gaensickeetal99-1}
{G\"ansicke}, B.~T., {Sion}, E.~M., {Beuermann}, K., {Fabian}, D., {Cheng},
  F.~H., {Krautter}, J., 1999, A\&A, 347, 178

\bibitem[{{Hamada} \& {Salpeter}(1961)}]{hamada+salpeter61-1}
{Hamada}, T., {Salpeter}, E.~E., 1961, ApJ, 134, 683

\bibitem[{{Hameury} \& {Lasota}(2005)}]{hameury+lasota05-1}
{Hameury}, J.-M., {Lasota}, J.-P., eds., 2005, The Astrophysics of Cataclysmic
  Variables and Related Objects, ASP Conf. Ser. 330

\bibitem[{{Howell} et~al.(1997){Howell}, {Rappaport}, \&
  {Politano}}]{howelletal97-1}
{Howell}, S.~B., {Rappaport}, S., {Politano}, M., 1997, MNRAS, 287, 929

\bibitem[{{Hubeny} \& {Lanz}(1995)}]{hubeny+lanz95-1}
{Hubeny}, I., {Lanz}, T., 1995, ApJ, 439, 875

\bibitem[{{Kirkpatrick} et~al.(1999)}]{kirkpatricketal99-1}
{Kirkpatrick}, J.~D., et~al., 1999, ApJ, 519, 802

\bibitem[{{Kirkpatrick} et~al.(2000)}]{kirkpatricketal00-1}
{Kirkpatrick}, J.~D., et~al., 2000, AJ, 120, 447

\bibitem[{{Koester} et~al.(1979){Koester}, {Schulz}, \&
  {Weidemann}}]{koesteretal79-1}
{Koester}, D., {Schulz}, H., {Weidemann}, V., 1979, A\&A, 76, 262

\bibitem[{{Kolb}(1993)}]{kolb93-1}
{Kolb}, U., 1993, A\&A, 271, 149

\bibitem[{{Liebert} et~al.(2005){Liebert}, {Bergeron}, \&
  {Holberg}}]{liebertetal05-1}
{Liebert}, J., {Bergeron}, P., {Holberg}, J.~B., 2005, ApJS, 156, 47

\bibitem[{{Mukadam} et~al.(2004)}]{mukadametal04-1}
{Mukadam}, A.~S., et~al., 2004, ApJ, 607, 982

\bibitem[{{Mukai} \& {Patterson}(2004)}]{mukai+patterson04-1}
{Mukai}, K., {Patterson}, J., 2004, in {Tovmassian}, G., {Sion}, E., eds.,
  Compact Binaries and Beyond, no.~20 in Conf. Ser., RMAA, p. 244

\bibitem[{{Mullally} et~al.(2005){Mullally}, {Thompson}, {Castanheira},
  {Winget}, {Kepler}, {Eisenstein}, {Kleinman}, \& {Nitta}}]{mullallyetal05-1}
{Mullally}, F., {Thompson}, S.~E., {Castanheira}, B.~G., {Winget}, D.~E.,
  {Kepler}, S.~O., {Eisenstein}, D.~J., {Kleinman}, S.~J., {Nitta}, A., 2005,
  ApJ, 625, 966

\bibitem[{{Patterson} et~al.(2005{\natexlab{a}}){Patterson}, {Thorstensen},
  {Armstrong}, {Henden}, \& {Hynes}}]{pattersonetal05-1}
{Patterson}, J., {Thorstensen}, J., {Armstrong}, E., {Henden}, A., {Hynes}, R.,
  2005{\natexlab{a}}, PASP, in press (astro-ph/0506135)

\bibitem[{{Patterson} et~al.(2005{\natexlab{b}}){Patterson}, {Thorstensen}, \&
  {Kemp}}]{pattersonetal05-2}
{Patterson}, J., {Thorstensen}, J.~R., {Kemp}, J., 2005{\natexlab{b}}, PASP,
  117, 427

\bibitem[{{Scargle}(1982)}]{scargle82-1}
{Scargle}, J.~D., 1982, ApJ, 263, 835

\bibitem[{{Schneider} \& {Young}(1980)}]{schneider+young80-1}
{Schneider}, D.~P., {Young}, P., 1980, ApJ, 240, 871

\bibitem[{{Schwarz} et~al.(2004){Schwarz}, {Barman}, {Silvestri}, {Szkody},
  {Starrfield}, {Vanlandingham}, \& {Wagner}}]{schwarzetal04-1}
{Schwarz}, G.~J., {Barman}, T., {Silvestri}, N., {Szkody}, P., {Starrfield},
  S., {Vanlandingham}, K., {Wagner}, R.~M., 2004, PASP, 116, 1111

\bibitem[{{Szkody} et~al.(2000){Szkody}, {Desai}, \& {Hoard}}]{szkodyetal00-1}
{Szkody}, P., {Desai}, V., {Hoard}, D.~W., 2000, AJ, 119, 365

\bibitem[{{Szkody} et~al.(2002{\natexlab{a}}){Szkody}, {G{\" a}nsicke},
  {Howell}, \& {Sion}}]{szkodyetal02-4}
{Szkody}, P., {G{\" a}nsicke}, B.~T., {Howell}, S.~B., {Sion}, E.~M.,
  2002{\natexlab{a}}, ApJ Lett., 575, L79

\bibitem[{{Szkody} et~al.(2002{\natexlab{b}})}]{szkodyetal02-2}
{Szkody}, P., et~al., 2002{\natexlab{b}}, AJ, 123, 430

\bibitem[{{Szkody} et~al.(2003)}]{szkodyetal03-2}
{Szkody}, P., et~al., 2003, AJ, 126, 1499

\bibitem[{{Szkody} et~al.(2004)}]{szkodyetal04-1}
{Szkody}, P., et~al., 2004, AJ, 128, 1882

\bibitem[{{Szkody} et~al.(2005)}]{szkodyetal05-1}
{Szkody}, P., et~al., 2005, AJ, 129, 2386

\bibitem[{{Thorstensen}(2003)}]{thorstensen03-1}
{Thorstensen}, J.~R., 2003, AJ, 126, 3017

\bibitem[{{Thorstensen} et~al.(2002){Thorstensen}, {Patterson}, {Kemp}, \&
  {Vennes}}]{thorstensenetal02-3}
{Thorstensen}, J.~R., {Patterson}, J., {Kemp}, J., {Vennes}, S., 2002, PASP,
  114, 1108

\bibitem[{{Tovmassian} et~al.(2003)}]{tovmassianetal03-1}
{Tovmassian}, G., et~al., 2003, PASP, 115, 725

\bibitem[{{Townsley} \& {Bildsten}(2003)}]{townsley+bildsten03-1}
{Townsley}, D.~M., {Bildsten}, L., 2003, ApJ Lett., 596, L227

\bibitem[{{Townsley} et~al.(2004){Townsley}, {Arras}, \&
  {Bildsten}}]{townsleyetal04-1}
{Townsley}, D.~M., {Arras}, P., {Bildsten}, L., 2004, ApJ Lett., 608, L105

\bibitem[{{van Zyl} et~al.(2004)}]{vanzyletal04-1}
{van Zyl}, L., et~al., 2004, MNRAS, 350, 307

\bibitem[{{Vanlandingham} et~al.(2005){Vanlandingham}, {Schwarz}, \&
  {Howell}}]{vanlandinghametal05-1}
{Vanlandingham}, K.~M., {Schwarz}, G.~J., {Howell}, S.~B., 2005, pasp, in press
  (astro-ph/0506098)

\bibitem[{{Voges} et~al.(2000)}]{vogesetal00-1}
{Voges}, W., et~al., 2000, IAU Circ., 7432

\bibitem[{{Warner} \& {Woudt}(2004)}]{warner+woudt04-1}
{Warner}, B., {Woudt}, P., 2004, in {Kurtz}, D.~W., {Pollard}, K.~R., eds.,
  Variable Stars in the Local Group, ASP Conf. Ser. 310, p. 392

\bibitem[{{Watson} et~al.(1996){Watson}, {Marsh}, {Fender}, {Barstow}, {Still},
  {Page}, {Dhillon}, \& {Beardmore}}]{watsonetal96-1}
{Watson}, M.~G., {Marsh}, T.~R., {Fender}, R.~P., {Barstow}, M.~A., {Still},
  M., {Page}, M., {Dhillon}, V.~S., {Beardmore}, A.~P., 1996, MNRAS, 281, 1016

\bibitem[{{Wheatley} et~al.(2000){Wheatley}, {Burleigh}, \&
  {Watson}}]{wheatleyetal00-1}
{Wheatley}, P.~J., {Burleigh}, M.~R., {Watson}, M.~G., 2000, MNRAS, 317, 343

\bibitem[{{Winget}(1998)}]{winget98-1}
{Winget}, D.~E., 1998, Journal of the Physics of Condensed Matter, 10, 11247

\bibitem[{{Woudt} \& {Warner}(2002)}]{woudt+warner02-2}
{Woudt}, P.~A., {Warner}, B., 2002, Ap\&SS, 282, 433

\bibitem[{{Woudt} \& {Warner}(2004)}]{woudt+warner04-1}
{Woudt}, P.~A., {Warner}, B., 2004, MNRAS, 348, 599

\bibitem[{{Woudt} et~al.(2005){Woudt}, {Warner}, {Pretorius}, \&
  {Dale}}]{woudtetal05-1}
{Woudt}, P.~A., {Warner}, B., {Pretorius}, M.~L., {Dale}, D., 2005, in
  \cite{hameury+lasota05-1}, p. 325, p. 325

\bibitem[{{Yaron} et~al.(2005){Yaron}, {Prialnik}, {Shara}, \&
  {Kovetz}}]{yaronetal05-1}
{Yaron}, O., {Prialnik}, D., {Shara}, M.~M., {Kovetz}, A., 2005, ApJ, 623, 398

\bibitem[{{Zharikov} et~al.(2005){Zharikov}, {Tovmassian}, {Neustroev},
  {Michel}, \& {Napiwotzki}}]{zharikovetal05-1}
{Zharikov}, S.~V., {Tovmassian}, G.~H., {Neustroev}, V., {Michel}, R.,
  {Napiwotzki}, R., 2005, in  \cite{hameury+lasota05-1}, p. 327, p. 327

\end{thebibliography}

\bsp

\label{lastpage}

\end{document}